\documentclass[review]{elsarticle}
\usepackage{amssymb}
\usepackage{graphicx}

\usepackage{lineno,hyperref}
\modulolinenumbers[5]

\journal{Physica B: Condensed Matter}

\begin{document}

\begin{frontmatter}

\title{Variational Monte Carlo simulation in hole-doped cuprate superconductors: Competition between antiferromagnetism and superconductivity}

\author{Chung-Pin Chou}
\address{Beijing Computational Science Research Center, Beijing 100084, China}

\begin{abstract}
We present variational Monte Carlo (VMC) results for the
Gutzwiller-projected coexisting state including both
antiferromagnetic (AFM) order and superconducting (SC) order in the
two-dimensional $t-t'-t''-J$ model. By further considering off-site
spin correlation between electrons, in contrast to earlier VMC
results [Phys. Rev. Lett. \textbf{102}, 027002 (2009)], we find the
apparent competition between AFM order and SC order near the
underdoped regime instead of coexistence. The local ferromagnetic
correlation introduced by spin-spin Jastrow correlators seem to be
responsible for the disappearance of the coexisting state. We also
demonstrate that the quasiparticle spectral weight from upper
(lower) AFM band are strongly diminished (enhanced) by the spin-spin
correlation. This result obviously leads to the loss of antinodal
electron pockets and the appearance of nodal hole pockets as passing
from the AFM phase to the SC phase in hole-doped cuprates, which is
in consistent with the observation by angle-resolved photoemission
spectroscopy.
\end{abstract}


\end{frontmatter}

%%%%%%%%%%%%%%%%%%%%%%%%%%%%%%%%%%%%%%%%%%%%%%%%%%%%%%%%%%%%%%%%%%%
\section{Introduction}
%%%%%%%%%%%%%%%%%%%%%%%%%%%%%%%%%%%%%%%%%%%%%%%%%%%%%%%%%%%%%%%%%%%
The doping phase diagram near the underdoped region is one of the
important and long-debated issues with high-$T_{c}$ cuprates
\cite{LeeRMP06}.
Since the parent compound is antiferromagnetic (AFM) Mott insulator,
the AFM correlation plays a significant role in the emergence of
superconductivity by doping charge carries.
The intrinsic proximity of the superconducting (SC) phase with the
AFM phase is also shared by the phase diagrams of other SC
materials, such as iron pnictides \cite{StewartRMP11} and heavy
fermion superconductors \cite{NicklasPRB07}.
The multi-layered cuprate superconductors exhibit the coexistence of
AFM and SC states at underdoping discovered by nuclear magnetic
resonance measurements \cite{MukudaJPSJ08,MukudaJPSJ09}.
However, the AFM phase and the SC phase never coexist in the phase
diagram of single-layered cuprates such as
La$_{2-x}$Sr$_{x}$CuO$_{4}$ \cite{KeimerPRB92} and
Bi$_{2}$Sr$_{2}$CuO$_{6+\delta}$ \cite{KatoJSSC97}.
In particular, Bi$_{2}$(Sr$_{2-x}$La$_{x}$)CuO$_{6+\delta}$ systems
shows that the three-dimensional AFM region, separated by the SC
phase, even survives until a high underdoping level
\cite{KawasakiPRL10}.

The existence of the coexisting state has been found by analytical
and numerical approaches in Hubbard$-$type models
\cite{ReissPRB07,JarrellEPL01,AichhornPRB07,DSPRL05,KobayashiPhysicaC10,CaponePRB06,KancharlaPRB08}
and $t-J-$type models
\cite{InabaPhysicaC96,HimedaPRB99,YamasePRB04,ShihPRB04,ShihLTP05,PathakPRL09,WatanabePhysicaC10}.
They seem to contribute the underlying mechanism to the coexisting
state observed in multi-layered cuprates.
However, a proper mechanism to explain why these two phases do not
like to coexist in single-layered cuprates remains needed.
Interestingly, some previous studies proposed the spin-bag mechanism
for superconductivity since two spin bags would attract each other
to form a Cooper pair and lower the total energy
\cite{SchriefferPRB89,WengPRB90,EderPRB94}.
As for doping more holes, therefore, it is necessary to re-examine
how the local distortion of the AFM background around holes
influences AFM order and SC order.

On the other hand, one of the most exciting experimental results is
the observation of quantum oscillations in the hole-doped cuprates
which pointed to electron pockets \cite{LeBoeufNat07,LeBoeufPRB11}.
In particular, they proposed that these electron pockets probably
originate from the Fermi surface reconstruction caused by the onset
of a density-wave phase, e.g. the AFM phase.
Unfortunately the electron-like Fermi pockets have never been found
in most of hole-doped cuprates using angle-resolved photoemission
spectroscopy (ARPES) \cite{YangNat08,MengNat09,LuARCMP12}.
Thus, to comprehend the loss of the electron pocket observed by
ARPES experiments, we inquire to what extent into the electronic
correlations ignored in mean-field calculations.

In this work, we study Gutzwiller's trial wave functions with the
coexistence of AFM order and SC order by means of variational Monte
Carlo (VMC) method.
To improve the trial state, we further consider the off-site
correlations between two electrons by applying suitable Jastrow
correlators.
Surprisingly, the long-range AFM order is strongly enhanced due to
the local ferromagnetic (FM) Jastrow correlation, or precisely local
AFM distortion, giving rise to the disappearance of the coexisting
state near underdoping in the phase diagram.
Besides, the spin-spin correlation in the non-coexisting state
transfers the quasiparticle spectral weight from the antinodal
electron pockets to the lower AFM band, and also the nodal hole
pockets can remain until superconductivity occurs.
Therefore, it is expected that the signal of the electron pockets
around antinodes cannot be found in many hole-doped compounds by
using ARPES.
%

%%%%%%%%%%%%%%%%%%%%%%%%%%%%%%%%%%%%%%%%%%%%%%%%%%%%%%%%%%%%%%%%%%%
\section{Theory}
%%%%%%%%%%%%%%%%%%%%%%%%%%%%%%%%%%%%%%%%%%%%%%%%%%%%%%%%%%%%%%%%%%%
Let us begin by the Hamiltonian on a square lattice of size
$16\times16$,
\begin{eqnarray}
H=-\sum_{i,j,\sigma}t_{ij}\tilde{c}_{i\sigma}^{\dag}\tilde{c}_{j\sigma}+J\sum_{\langle
i,j\rangle}\left(\mathbf{S}_{i}\cdot\mathbf{S}_{j}-\frac{1}{4}n_{i}n_{j}\right),
\label{e:equ1}
\end{eqnarray}
where the hopping $t_{ij}=t$, $t'$, and $t''$ for sites i and j
being the nearest, second-nearest, and third-nearest neighbors,
respectively.
Other notations are standard. We restrict the electron creation
operators $\tilde{c}_{i\sigma}^{\dag}$ to the subspace without
doubly-occupied sites.
In the following, the bare parameters $(t',t'',J)/t$ in the
Hamiltonian are set to be in the hole-doped regime:
$(-0.3,0.15,0.3)$.
In order to understand how AFM order and SC order compete in
variational phase diagram, we choose the mean-field ground state
including both AFM order and SC order (AFSC) as a starting point,
\begin{eqnarray}
|\Psi_{AFSC}\rangle=\prod'_{\bf{k},s=\{a,b\}}\gamma_{\bf{k}\uparrow}^{s}\gamma_{-\bf{k}\downarrow}^{s}|0\rangle,\label{e:equ2}
\end{eqnarray}
where the prime means the product only includes momenta inside the
magnetic zone boundary (MZB).
Note that $s$ represents the quasiparticle coming from the upper AFM
band ($s=b$) or the lower AFM band ($s=a$).
The Bogoliubov's quasiparticle operators $\gamma_{\bf{k}\sigma}^{s}$
are defined as
\begin{eqnarray}
\gamma_{\bf{k}\sigma}^{s}=u_{\bf{k}}^{s}\hat{s}_{\bf{k}\sigma}-\sigma
v_{\bf{k}}^{s}\hat{s}_{-\bf{k}\bar{\sigma}}^{\dag}.\label{e:equ5}
\end{eqnarray}
The coefficients $u_{\bf{k}}^{s}$ and $v_{\bf{k}}^{s}$ are the BCS
coherence factor of AFM quasiparticles corresponding to the $s$
band,
\begin{eqnarray}
(u_{\bf{k}}^{s})^{2}&=&\frac{1}{2}\left(1+\frac{\xi_{\bf{k}}^{s}}{\sqrt{(\xi_{\bf{k}}^{s})^{2}+\Delta_{\bf{k}}^{2}}}\right),\nonumber\\
(v_{\bf{k}}^{s})^{2}&=&1-(u_{\bf{k}}^{s})^{2},\label{e:equ6}
\end{eqnarray}
where the AFM band dispersion
$\xi_{\bf{k}}^{b/a}=\epsilon_{\bf{k}}^{+}\pm\sqrt{(\epsilon_{\bf{k}}^{-})^{2}+m^{2}}$
and
$\epsilon_{\bf{k}}^{\pm}\equiv\left(\varepsilon_{\bf{k}}\pm\varepsilon_{\bf{k+Q}}\right)/2$.
Here $\varepsilon_{\bf{k}}$ is the normal-state dispersion.
$\Delta_{\bf{k}}$($=2\Delta\left(\cos\bf{k_{x}}-\cos\bf{k_{y}}\right)$)
is $d$-wave pairing amplitude and $m$ AFM order parameter.
The annihilation operators for AFM bands, $\hat{s}_{\bf{k}\sigma}$,
are given by
\begin{eqnarray}
\left(
  \begin{array}{c}
    a_{\bf{k}\sigma} \\
    b_{\bf{k}\sigma} \\
  \end{array}
  \right)=\left(
          \begin{array}{cc}
            \alpha_{\bf{k}} & \sigma\beta_{\bf{k}} \\
            -\sigma\beta_{\bf{k}} & \alpha_{\bf{k}} \\
          \end{array}
        \right)\left(
                 \begin{array}{c}
                   c_{\bf{k}\sigma} \\
                   c_{\bf{k+Q}\sigma} \\
                 \end{array}
               \right),\label{e:equ3}
\end{eqnarray}
with $\bf{Q}=(\pi,\pi)$ and the coefficients
\begin{eqnarray}
\alpha_{\bf{k}}^{2}&=&\frac{1}{2}\left(1-\frac{\epsilon_{\bf{k}}^{-}}{\sqrt{(\epsilon_{\bf{k}}^{-})^{2}+m^{2}}}\right),\nonumber\\
\beta_{\bf{k}}^{2}&=&1-\alpha_{\bf{k}}^{2}.\label{e:equ4}
\end{eqnarray}

In order to introduce more correlations in the mean-field wave
function, we first formulate the trial wave function fixing the
number of electrons $\hat{P}_{N_{e}}$ with on-site Gutzwiller
projector
$\hat{P}_{G}(=\prod_{i}\left(1-\hat{n}_{i\uparrow}\hat{n}_{i\downarrow}\right))$
and charge-charge Jastrow correlator ($\hat{P}_{J}^{CC}$)
\cite{CPC08,CPC12},
\begin{eqnarray}
|\Psi_{CC}\rangle=
\hat{P}_{N_{e}}\hat{P}_{G}\hat{P}_{J}^{CC}|\Psi_{AFSC}\rangle.\label{e:equ7}
\end{eqnarray}
More importantly, we also consider the correlation between spins by
using spin-spin Jastrow correlator ($\hat{P}_{J}^{SS}$),
\begin{eqnarray}
|\Psi_{CCSS}\rangle=\hat{P}_{J}^{SS}|\Psi_{CC}\rangle.\label{e:equ82}
\end{eqnarray}
The Jastrow correlator is constructed by classical Boltzmann
operator, $\hat{P}_{J}^{i}=e^{\hat{H}_{i}}$, encoding the intersite
correlations.
For the sake of simplicity, $\hat{H}_{i}$ depicting charge ($i=CC$)
and spin ($i=SS$) parts are chosen to be diagonal in real-space
configuration.
The charge-charge Jastrow correlator describes the short- and
long-range correlations between holes in the lattice system.
Thus,
\begin{eqnarray}
\hat{H}_{CC}=\sum_{i<j}\eta_{ij}\hat{n}_{i}^{h}\hat{n}_{j}^{h},\label{e:equ81}
\end{eqnarray}
with
$\eta_{ij}\equiv\ln(r_{ij}^{\alpha}v_{\gamma}^{\delta_{j,i+\gamma}})$.
Here $r_{ij}$ is the chord length of $|\vec{r}_{i}-\vec{r}_{j}|$ and
$\hat{n}_{i}^{h}=1-\sum_{\sigma}\hat{n}_{i\sigma}$.
We consider three parameters $v_{\gamma}$, the nearest ($\gamma=1$),
second-nearest ($\gamma=2$) and third-nearest ($\gamma=3$)
neighbors, standing for short-range hole-hole repulsion if
$v_{\gamma}<1$.
The factor $r_{ij}^{\alpha}$ denotes attractive long-range
($r_{ij}>1$) and repulsive short-range ($r_{ij}<1$) correlations
between holes if $\alpha>0$.

A similar formalism to the spin-spin correlation has been considered
at half-filling \cite{HuesPRL88}.
We further imitate the formalism described above to write down the
spin-spin Jastrow correlator,
\begin{eqnarray}
\hat{H}_{SS}=\sum_{i<j}\kappa_{ij}\hat{S}_{i}^{z}\hat{S}_{j}^{z},\label{e:equ8}
\end{eqnarray}
where
$\kappa_{ij}\equiv\ln(r_{ij}^{\beta}w_{\gamma}^{\delta_{j,i+\gamma}})$
and $\hat{S}_{i}^{z}$ the spin operator along $z$ direction at site
$i$.
The only difference from the charge counterpart is that the
sign of $\hat{S}_{i}^{z}\hat{S}_{j}^{z}$ determines the type of
magnetic correlations.
In other words, it will be the FM (AFM) correlation if
$\hat{S}_{i}^{z}\hat{S}_{j}^{z}>0$ ($<0$).
In addition to the parameter $\beta$ controlling the long-range spin
correlations, we consider the other three parameters
$w_{\gamma=1,2,3}$ for the neighboring spin-spin correlations.
For example of the FM case, the short-range correlation would be
suppressed when $w_{\gamma}<1$.
On the other hand, the factor $r_{ij}^{\beta}$ control the
long-range ($r_{ij}>1$) and short-range ($r_{ij}<1$) correlations.
In the long-range case of $\beta<0$, for instance, $r_{ij}^{\beta}$
would decrease the FM correlation but conversely increase the AFM
correlation.

In addition to the ground state, we also propose a trial wave
function for the low-lying excitation of the Gutzwiller-projected
coexisting state simply generated by Gutzwiller projecting the
mean-field excited state
\begin{eqnarray}
|\Psi_{AFSC}^{\bf{k}\sigma
s}\rangle=\left(\gamma_{\bf{k}\sigma}^{s}\right)^{\dag}|\Psi_{AFSC}\rangle.\label{e:equ9}
\end{eqnarray}
Here we have applied the particle-hole transformation
\cite{YokoyamaJPSJ88,CPCPRB12} into Eq.(\ref{e:equ9}) to avoid the
divergence from the nodes of the mean-field wave function.
The Gutzwiller-projected excited state with both AFM order and SC
order fixing to $N_{e}-1$ electrons is written as
\begin{eqnarray}
|\Psi_{\bf{k}\sigma}^{s}\rangle=
\hat{P}_{N_{e}-1}\hat{P}_{G}\hat{P}_{J}^{CC}\hat{P}_{J}^{SS}|\Psi_{AFSC}^{\bf{k}\sigma
s}\rangle.\label{e:equ10}
\end{eqnarray}
Hence we can compute the excitation energies
$E_{k}(\equiv\langle\Psi_{\bf{k}\sigma}^{s}|H|\Psi_{\bf{k}\sigma}^{s}\rangle-\langle\Psi_{0}|H|\Psi_{0}\rangle)$
for either upper ($s=b$) or lower ($s=a$) AFM quasiparticles.
Furthermore, the quasiparticle spectral weight measured from ARPES
can be obtained by calculating
\begin{eqnarray}
Z_{\bf{k}}^{-}\equiv\frac{\left|\langle\Psi_{\bf{k}\sigma}^{s}|c_{-\bf{k}\bar{\sigma}}|\Psi_{0}\rangle\right|^{2}}{\langle\Psi_{\bf{k}\sigma}^{s}|\Psi_{\bf{k}\sigma}^{s}\rangle\langle\Psi_{0}|\Psi_{0}\rangle}.\label{e:equ11}
\end{eqnarray}
Some details in the VMC calculation should be noticed.
The boundary condition we use is periodic along both directions.
In order to achieve a reasonable acceptance ratio, the simulation
consists of a combination of one-particle moves and two-particle
moves.
The variational parameters of the Gutzwiller-projected coexisting
state are optimized by using the stochastic reconfiguration method
\cite{SorellaPRB01}.
All physical quantities are evaluated using the optimized
parameters.
We also take a sufficient number of samples ($=2\times10^{5}$) to
reduce the statistical errors, and keep the sampling interval
($\sim40$) long enough to ensure statistical independence between
samples.
%

%%%%%%%%%%%%%%%%%%%%%%%%%%%%%%%%%%%%%%%%%%%%%%%%%%%%%%%%%%%%%%%%%%%
\section{Results}
%%%%%%%%%%%%%%%%%%%%%%%%%%%%%%%%%%%%%%%%%%%%%%%%%%%%%%%%%%%%%%%%%%%

\begin{figure}[t]
\begin{center}\rotatebox{0}{\includegraphics[height=3.5in,width=3in]{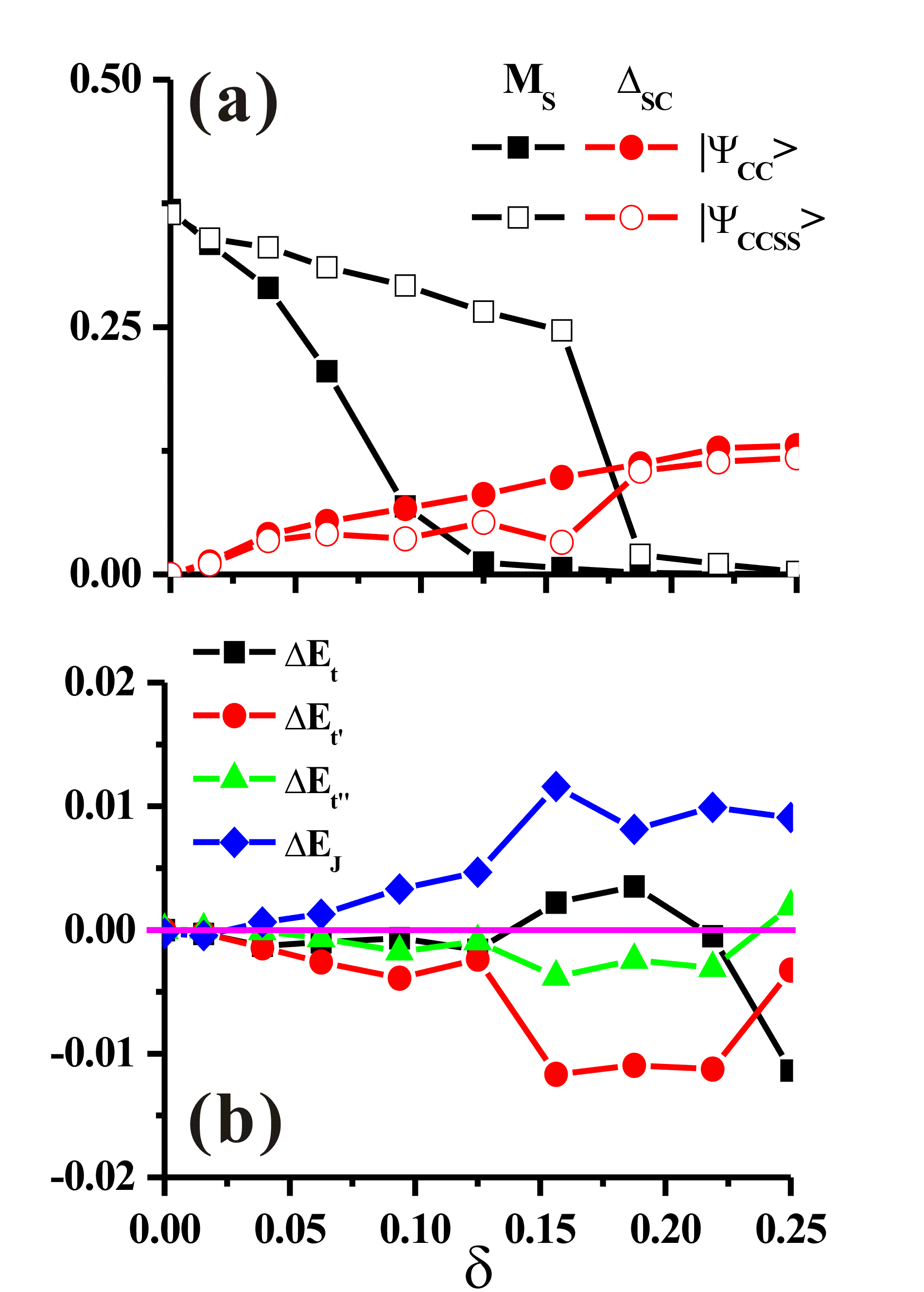}}\end{center}
\caption{(a) Variational phase diagram plotted by staggered
magnetization $M_{s}$ (squares) and superconducting order parameter
$\Delta_{SC}$ (circles). Filled and empty symbols represent
$|\Psi_{CC}\rangle$ and $|\Psi_{CCSS}\rangle$, respectively. (b) The
difference of the energy components between $|\Psi_{CCSS}\rangle$
and $|\Psi_{CC}\rangle$ as a function of hole doping $\delta$ in
$16\times16$ lattice.}\label{fig1}
\end{figure}

We first consider the trial state with only the charge-charge
Jastrow correlator to better demonstrate the variational phase
diagram.
Then we further include the spin-spin Jastrow correlator to
see how the phase diagram changes.
Order parameters shown in the phase diagram are determined by the
staggered magnetization
\begin{eqnarray}
M_{s}=\frac{1}{N}\sum_{i}\langle\hat{S}_{i}^{z}\rangle
e^{i\bf{Q}\cdot\bf{R}_{i}}\label{e:mag}
\end{eqnarray}
and the long-range pair-pair correlation function
\begin{eqnarray}
C_{PP}(R)=\frac{1}{N}\sum_{i,\alpha,\alpha'}\lambda_{\alpha,\alpha'}\langle\Delta_{i,\alpha}^{\dag}\Delta_{i+R,\alpha'}\rangle.\label{e:ppcf}
\end{eqnarray}
The creation operator
$\Delta_{i,\alpha}^{\dag}$($\equiv\tilde{c}_{i\uparrow}^{\dag}\tilde{c}_{i+\alpha\downarrow}^{\dag}-\tilde{c}_{i\downarrow}^{\dag}\tilde{c}_{i+\alpha\uparrow}^{\dag}$)
creates a singlet on the bond $(i,i+\alpha)$, $\alpha=x,y$.
The factor $\lambda_{\alpha,\alpha'}$ describes $d$-wave symmetry:
$\lambda_{\alpha,\alpha'}=1$($-1$) as
$\alpha=\alpha'$($\alpha\neq\alpha'$).

In Fig.\ref{fig1}(a), without the spin-spin Jastrow correlators as
indicated by $|\Psi_{CC}\rangle$, there exists a region showing the
coexistence of AFM order and SC order within doping
$\delta\lesssim0.125$ in the phase diagram
\cite{ShihLTP05,PathakPRL09,WatanabePhysicaC10}, where $M_{s}$ and
$\Delta_{SC}$($\equiv\sqrt{C_{PP}(R>2)}$) are finite.
Let us turn to the case with both charge-charge and spin-spin
Jastrow correlators denoted by $|\Psi_{CCSS}\rangle$.
Obviously the coexisting region disappears and a clear boundary
separating the AFM phase and the SC phase shows up at doping
$\delta=0.156$.
Note that near the boundary the spin-spin Jastrow correlator can
greatly improve the ground-state energies from $0.3\%$ to $0.7\%$.
From the numerical optimization, we find the spin-spin Jastrow
correlator can provide a conduit to vary the mean-field AFM order in
$|\Psi_{AFSC}\rangle$.
Surprisingly, the optimized spin-spin Jastrow parameters slightly
display short-range FM correlations in the AFM background (e.g. at
$\delta=0.156$ the spin-spin Jastrow weights $w_{\gamma}$ for
$\gamma=1$, $2$ and $3$ would be increased to $1.12$, $1.02$ and
$1.01$, respectively).
The local FM correlation introduced by the Jastrow factors is
harmful to the mean-field AFM order.
To make them balance, it is inevitable to largely enhance the AFM
background in $|\Psi_{AFSC}\rangle$.
The surprising competition between AFM order and SC order near the
phase boundary is mainly due to the hugely enhanced AFM order
further leading to the diminished SC order.

To further demonstrate the energy competition, we analyze the
difference of the energy components in the Hamiltonian between
$|\Psi_{CCSS}\rangle$ and $|\Psi_{CC}\rangle$ shown in
Fig.\ref{fig1}(b).
Our data clearly show that within $0.04<\delta<0.2$ the spin Jastrow
correlator helps the trial mean-field state gain much more energy
from the second-nearest-neighbor hopping term.
On the other hand, the competing energy primarily comes from the
spin-spin superexchange interaction.
From real-space point of view, holes prefer to move along diagonal
direction in strong AFM background so that the hopping energy from
the second nearest neighbors ($t'$) is likely to compete with the
superexchange energy ($J$).

\begin{figure}[t]
\begin{center}\rotatebox{0}{\includegraphics[height=2.4in,width=3.4in]{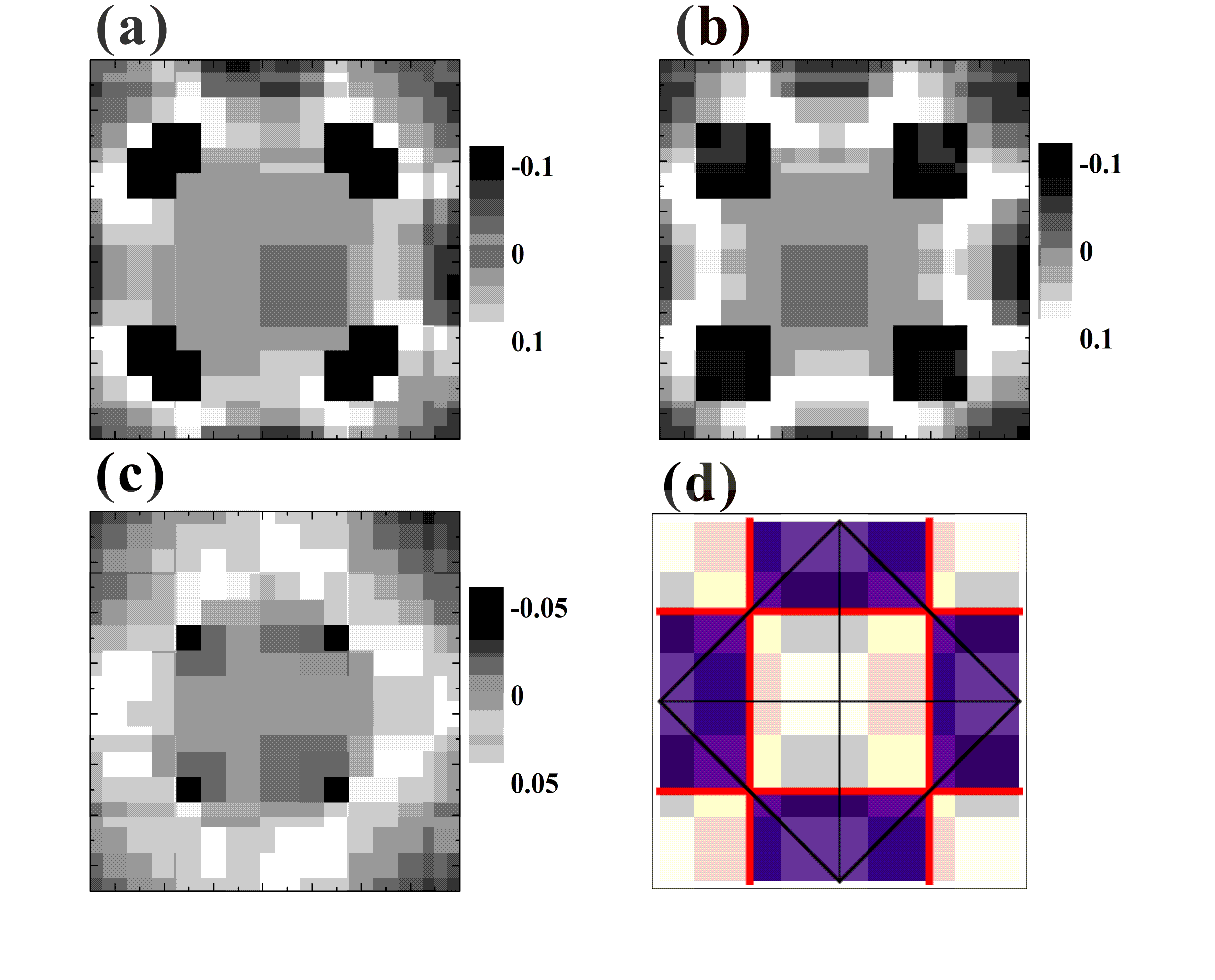}}\end{center}
\caption{The difference of the momentum distribution function
between $|\Psi_{CCSS}\rangle$ and $|\Psi_{CC}\rangle$ for doping (a)
$\delta=0.125$, (b) $\delta=0.156$ and (c) $\delta=0.188$ plotted in
the first Brillouin zone. (d) The next-nearest-neighbor energy
component, $-4t'\cos(k_{x})\cos(k_{y})$. The black diamond is the
half-filled Fermi surface. White (Purple) regions present the
positive (negative) values. Red lines mean zero.}\label{fig2}
\end{figure}

In momentum space, it is apparent that the $t'$ energy gain would
influence how the band dispersion evolves from Fermi pocket to Fermi
surface as increasing doping.
In Fig.\ref{fig2}(a)-(c), the difference of the momentum
distribution function between $|\Psi_{CCSS}\rangle$ and
$|\Psi_{CC}\rangle$ shows how electrons distribute in the band
structure.
At $\delta=0.125$ (Fig.\ref{fig2}(a)), obviously electrons in the
system would prefer to stay around "hot spots" rather than living
near nodes and antinodes, which hole pockets and electron pockets
seem to be observed as well.
The hot spot is defined as the momenta along the MZB that can be
connected by ($\pi,\pi$) momentum scattering.
Once doping is increased to $0.156$ which is the phase boundary
(Fig.\ref{fig2}(b)), hole pockets become larger and electron pockets
slightly shrink.
Now that electrons like to circle just outside the electron pockets,
they attempt to form a large Fermi surface.
Indeed, as further increasing doping to $0.188$ where the long-range
AFM order almost disappears (Fig.\ref{fig2}(c)), a clear Fermi
surface in which electrons cluster together can be seen.
So far, we also understand the reason why the system gain much
energy from $t'$ term since the hot spots are located right at the
purple region shown in Fig.\ref{fig2}(d).

\begin{figure}[t]
\begin{center}\rotatebox{0}{\includegraphics[height=3.7in,width=2.4in]{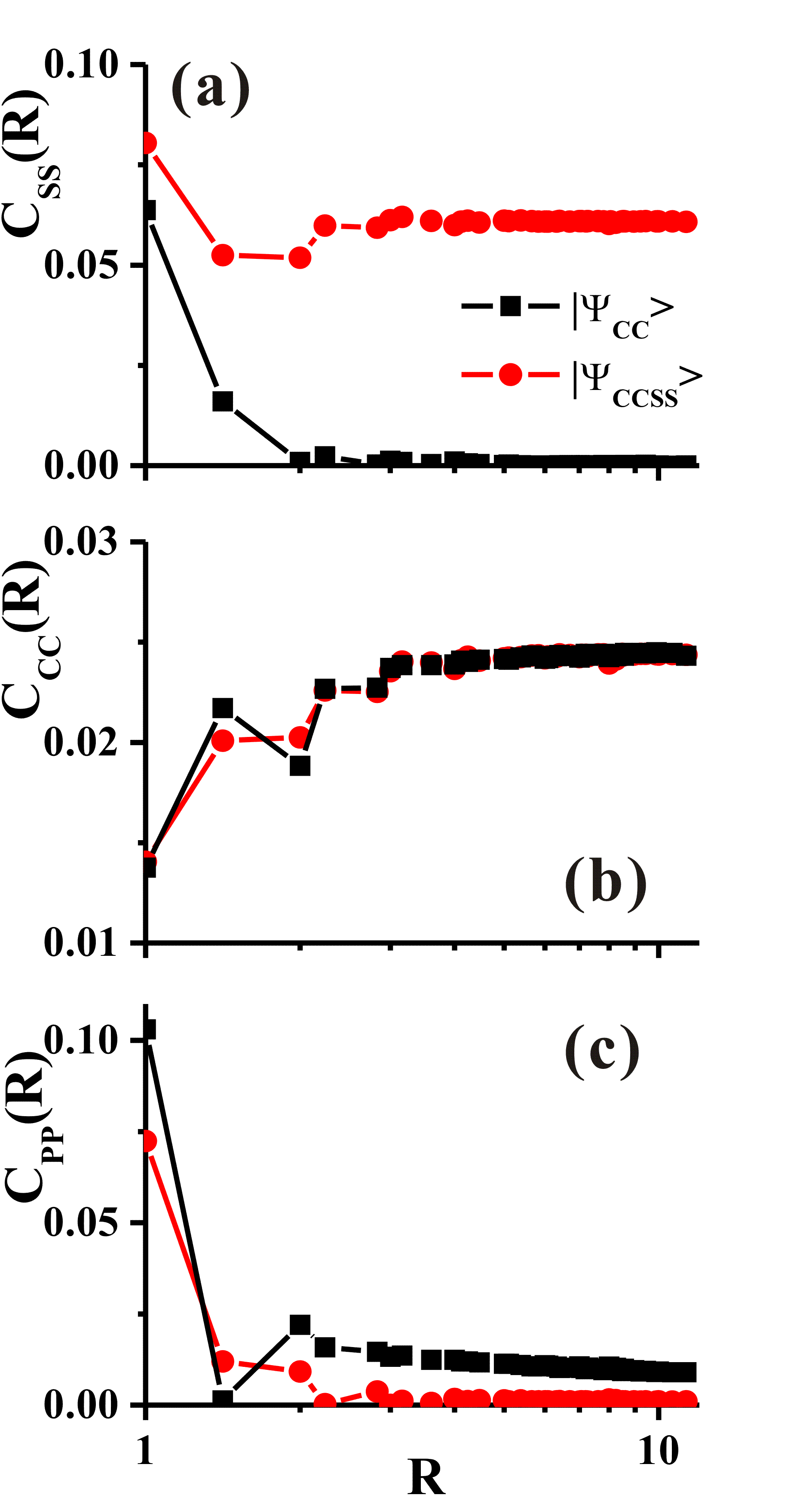}}\end{center}
\caption{(a) Spin-spin, (b) hole-hole and (c) pair-pair correlation
functions for the optimized state $|\Psi_{CCSS}\rangle$
($|\Psi_{CC}\rangle$), denoted by red circle (black square) symbols.
There are 40 doped holes in $16\times16$ lattice
($\delta=0.156$).}\label{fig3}
\end{figure}

In Fig.\ref{fig3}, we compute the spin-spin, hole-hole and pair-pair
correlation functions (already shown in Eq.(\ref{e:ppcf})) defined
as,
\begin{eqnarray}
C_{CC}(\bf{R})&=&\frac{1}{N}\sum_{i}\langle\hat{n}_{i}^{h}\hat{n}_{i+\bf{R}}^{h}\rangle,\\
C_{SS}(\bf{R})&=&\frac{1}{N}\sum_{i}\langle
\hat{S}_{i}^{z}\hat{S}_{i+\bf{R}}^{z}\rangle
e^{i\bf{Q}\cdot\bf{R}}.\label{e:equ12}
\end{eqnarray}
The doping density we choose to present is $0.156$.
Figure \ref{fig3}(a) illustrates that the spin-spin Jastrow
correlators indirectly induce the stronger AFM background showing a
constant tail in the staggered spin-spin correlation function which
implies a clear AFM order.
Note that The enhancement of the AFM order mainly arises from the
mean-field wave function $|\Psi_{AFSC}\rangle$.
Furthermore, we find in Fig.\ref{fig3}(b) that the hole-hole
correlation function makes no difference even if including the
spin-spin Jastrow correlators, except that the short-range part
becomes less staggered.
For spin and charge, there is no correlation for their long-range
behavior.
Finally, we can also see in Fig.\ref{fig3}(c) that as considering
$\hat{P}_{J}^{SS}$ the pair-pair correlation almost vanishes at
large distances so that the SC properties is not available.

\begin{figure}[t]
\begin{center}\rotatebox{0}{\includegraphics[height=2in,width=3.4in]{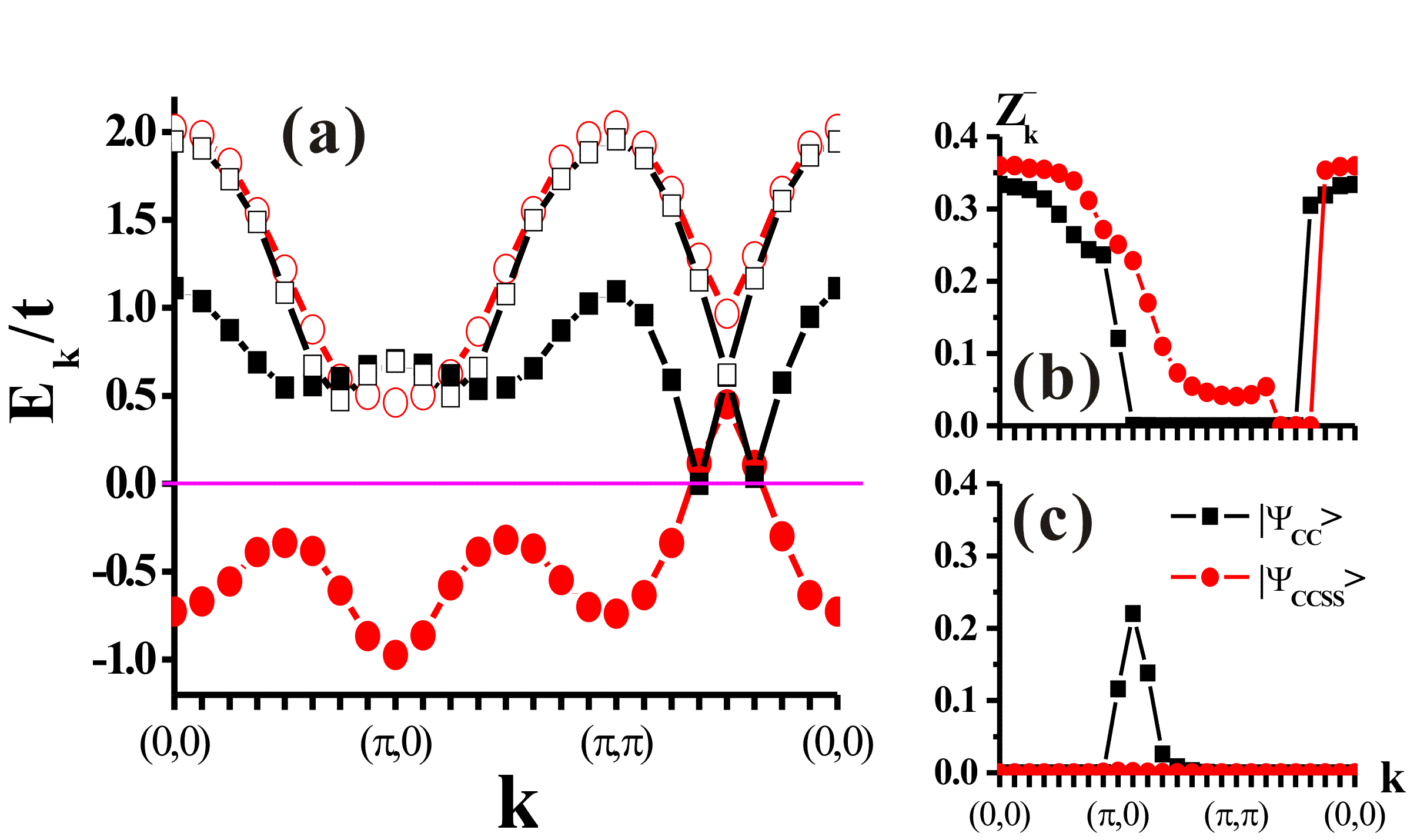}}\end{center}
\caption{(a) The quasi-particle excitation dispersion $E_{\bf{k}}$
for different optimized states (denoted in the legend of (c)) along
high symmetric momenta at $\delta=0.156$. Empty (Filled) symbols
represent the upper (lower) AFM band and squares (circles) the trial
state $|\Psi_{CC}\rangle$ ($|\Psi_{CCSS}\rangle$). Due to much
smaller $\Delta$ than $m$ for the trial state $|\Psi_{CCSS}\rangle$,
we simply plot the lower AFM band (red circles) below the Fermi
level (pink line) except the nodal regions for clear demonstration.
The quasiparticle spectral weight $Z_{\bf{k}}^{-}$ are obtained from
(b) the lower AFM band and (c) the upper AFM band.}\label{fig4}
\end{figure}

Next, it would be interesting to examine the low-lying
single-particle excitation spectra near the phase boundary.
In Fig.\ref{fig4}, by applying the ansatz (Eq.(\ref{e:equ10})) to
the single-particle excitation, we calculate two quasiparticle band
dispersions ($s=a,b$) and their corresponding spectral weight for
removing one particle defined by Eq.(\ref{e:equ11}).
In order to compare with the excitations with/without spin-spin
Jastrow correlators $\hat{P}_{J}^{SS}$, we plot their excitation
energy $E_{\bf{k}}$ along the high symmetric momenta in
Fig.\ref{fig4}(a).
In the case where the trial state only includes the charge-charge
Jastrow factors, its optimized mean-field parameters $\Delta\gg m$.
Due to large $d$-wave BCS pairing contribution, the dispersions thus
show convex around the antinodes and almost zero gap between the two
bands at nodes.
Especially, the upper AFM band is beneath the lower AFM band near
the antinodal regions, and hence there is a clear signal of electron
pockets arising from the upper AFM band shown in Fig.\ref{fig4}(c).

When further considering spin-spin Jastrow correlators, the
optimized mean-field parameters $m\gg\Delta$.
Such a huge AFM parameter $m$ gives rise to a typical AFM band
dispersion and opens a AFM gap between these two bands at nodes, as
indicated by red circles in Fig.\ref{fig4}(a).
Interestingly, Fig.\ref{fig4}(c) shows that near antinodes the
quasiparticle spectral weight of the upper AFM band disappear and
transfer to almost entire lower AFM band (see Fig.\ref{fig4}(b)).
In particular, a clear hole pocket of the lower AFM band centering
around $\bf{Q}/2$ is also observed in Fig.\ref{fig4}(b).
The Gutzwiller and Jastrow correlators arising from electronic
correlation firmly influence the low-lying quasiparticle excitation
spectra of the mean-field state $|\Psi_{AFSC}\rangle$.
Therefore, the loss of the electron pockets due to electron
correlations provides a route to figure out why electron pockets
have never been found in most of hole-doped cuprates measured by
ARPES.
%

%%%%%%%%%%%%%%%%%%%%%%%%%%%%%%%%%%%%%%%%%%%%%%%%%%%%%%%%%%%%%%%%%%%
\section{Conclusions}
%%%%%%%%%%%%%%%%%%%%%%%%%%%%%%%%%%%%%%%%%%%%%%%%%%%%%%%%%%%%%%%%%%%
Summing up, by using VMC approach we have studied the coexisting
state with both AFM order and SC order simultaneously underneath the
Gutzwiller's projection and Jastrow correlators.
We have thereby re-examined the variational ground-state phase
diagram and found that the AFM phase competes with the SC phase as
further considering off-site spin correlations.
The reasoning for the competition is that the mean-field AFM order
is considerably enhanced due to short-range FM correlation
introduced by the spin-spin Jastrow factors, further leading to the
vanished SC order.
As well, we have first investigated the Gutzwiller-projected
quasiparticle excitations of the coexisting state.
Based on the Gutzwiller ansatz, passing through the boundary between
AFM and SC phases, we have observed the loss of electron pockets
near antinodes coming from the upper AFM band and the occurrence of
hole pockets near nodes arising from the lower AFM band as long as
the spin-spin Jastrow correlators are included.
Therefore, such a strongly correlated electron system needs to be
carefully inspected in the explanation for the low-lying
quasiparticle excitations observed by ARPES experiments.

\section{Acknowledgments}
\label{Acknowledgment} Greatly thanks S.-M. Huang, W. Ku and T.-K.
Lee for helpful discussions. This work is supported by the
Postdoctoral Research Abroad Program sponsored by National Science
Council in Taiwan with Grant No. NSC 101-2917-I-564-010 and by CAEP
and MST. All calculations are performed in the National Center for
High-performance Computing in Taiwan.
\\

\end{document}